# Dielectric Resonator Antenna Coupled Antimonide-Based Detectors (DRACAD) For the Infrared

Jordan Budhu, *Member, IEEE*, Nicole Pfiester, Kwong-Kit Choi, *Fellow, IEEE*, Steve Young, Chris Ball, *Member, IEEE*, Sanjay Krishna, *Fellow, IEEE*, and Anthony Grbic, *Fellow, IEEE*

*Abstract*—In an infrared photodetector, noise current (dark current) is generated throughout the volume of the detector. Reducing the volume will reduce dark current, but the corresponding smaller area will also reduce the received signal. By using a separate antenna to receive light, one can reduce the detector area without reducing the signal, thereby increasing its signal to noise ratio (SNR). Here, we present a dielectric resonator antenna (DRA) coupled infrared photodetector. Its performance is compared to a conventional resonant cavity enhanced slab detector. The Noise Equivalent Power (NEP) is used as a figure of merit for the comparison. Formulas for the NEP are derived for both cases. A pBp photodiode detector is assumed in the comparison. The active region of the photodiode is an InAs/GaSb Type II Superlattice (T2SL). A Genetic Algorithm is used to optimize the dimensions of the detector and the DRA to achieve the smallest NEP. The result is a photodetector that converts over 85% of the incident light into carriers with a volume reduced by 95%. This optimal geometry leads to a NEP reduced by 6.02dB over that of the conventional resonant cavity enhanced slab detector.

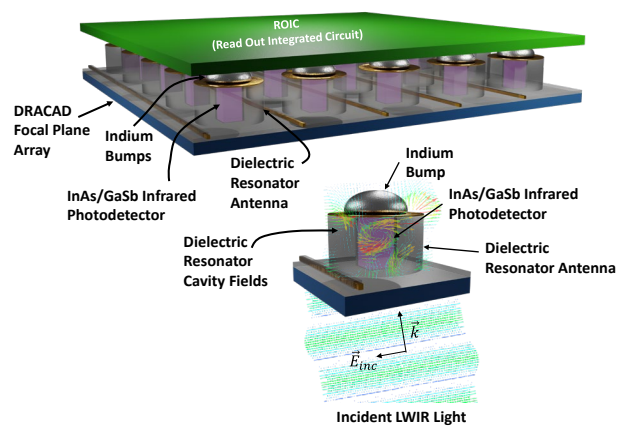

Fig 1. Dielectric Resonator Antenna Coupled Antimonide-Based Detector (DRACAD) focal plane array. The DRACAD is interfaced to the Read Out Integrated Circuit (ROIC) through indium bumps deposited over the gold caps of the DRACAD.

## I. Introduction

THE low bandgap energy associated with Long Wave Infrared (LWIR) photon detection leads to large dark currents. At elevated temperatures, carriers can easily gain sufficient thermal energy to cross the bandgap and generate electron-hole pairs which are read as signal even in the absence of incident radiation. It is therefore necessary to cryogenically cool the detector to reduce its dark current. The electron-hole pairs associated with dark current are generated throughout the detector's active volume [1]. Thus, one can reduce the dark current by reducing its volume. This in turn reduces the collected optical signal. The solution is to couple the reduced size detector to an antenna that has a large effective area relative to the detector dimensions [2,3,4]. The noiseless antenna channels the incident photons over a large area into a small detector volume for absorption. The detector is thus able to achieve a large signal collection and a small dark current at the same time. It can either improve the signal to noise ratio or raise the operating temperature.

An added benefit of the antenna coupled configuration is that the antenna acts as an impedance matching network between free space and the detector material. Since semiconducting materials typically have large indices of refraction, the mismatch at the air-semiconductor interface leads to substantial reflection loss, limiting the maximum achievable efficiency. The antenna coupled configuration reduces this reflection loss by providing a matched impedance to the incident wave.

In literature, there are many examples of antenna coupled infrared detectors. However, they are generally not applicable to densely populated 2-dimensional detector arrays, and they do not address the Antimonide-based detectors used in this work

Original manuscript received November 2020. This work was supported in part by the Army Research Office under grant no. W911NF-19-1-0359.

J. Budhu, S. Young, and A. Grbic are with the Department of Electrical Engineering and Computer Science, University of Michigan, Ann Arbor, MI 48109 USA (e-mail: jbudhu@umich.edu, yms@umich.edu, agrbic@umich.edu).

N. Pfiester and S. Krishna are with the Department of Electrical and Computer Engineering, The Ohio State University, Columbus, OH, USA 43210 (email: pfiesterlatham.1@osu.edu, krishna.53@osu.edu)

K.K. Choi is with Science Systems and Applications, Inc., Lanham, MD, USA 20706 (email: KNYCHOI@msn.com)

C. Ball is with the ElectroScience Laboratory, The Ohio State University, Columbus, OH, USA 43212 (email: ball.51@osu.edu)

specifically. The more common approach for 2-dimensional arrays is to use metamaterials for absorption enhancement. At LWIR frequencies, metals are plasmonic and exhibit appreciable conduction losses. In [5], losses in the metal layers total nearly 50% of the incident power. In [6], metal losses range from 21.91% to 36.30%. To avoid these losses, a Dielectric Resonator Antenna (DRA) made of amorphous silicon ($\alpha$-$Si$) is coupled to the detector. The choice of DRA avoids the losses associated with plasmonic metals, allows for high resolution focal plane arrays with high pixel count, and can achieve high total efficiencies. The dielectric resonator can also be made to resonate in different modes and thus allows for great design flexibility in its physical dimensions [7]. Furthermore, shaping the dielectric resonator antenna can lead to multiwavelength or broadband operation [8,9]. A focal plane array made with DRA coupled detector elements is shown schematically in Fig. 1.

One way to characterize a detector's sensitivity is through its Noise Equivalent Power (NEP). The NEP is defined as the amount of input power required to generate an output signal equal to the root-mean-square (RMS) of the noise generated by the detector in the absence of radiation [10]. Thus, the NEP is the input power which gives a Signal-to-Noise Ratio (SNR) of one at the output. Since the NEP is a measure of the minimum detectable power of the device, it is a good metric to quantify the detector's sensitivity and allows for a direct comparison of different detector configurations. In this work, the NEP is used to quantify the improvement of the antenna coupled configuration over that of the conventional detector made from a resonant cavity enhanced slab of absorbing material. Resonant cavity enhanced slab detectors are commonly used in infrared photodetectors and form a useful basis for comparison [11,12]. In the appendix, the NEP is precisely defined and derived. These formulations are used to calculate the SNR of the antenna coupled configuration and the conventional resonant cavity enhanced slab detector.

First, the photodetector material is presented along with its absorption coefficient curve. The material absorption characteristics are incorporated in the design of the dielectric resonator antenna through a Genetic Algorithm optimization. The total efficiency spectrum of the DRA coupled detector design is presented over the LWIR wavelength range. The SNR is calculated for our design and compared to the conventional detector's SNR. The DRA coupled detector is then compared to other state-of-the-art antenna couped photodetector designs. The total efficiency versus incident angle of the illuminating plane wave is reported. It is shown that the antenna coupled design is compatible with $f/2$ optics. In the Appendix A, a review of the theory of antenna coupled photodetectors is presented. In this appendix, expressions are derived and used to quantitatively compare the antenna coupled detector's noise performance to conventional detectors made of a resonant slab of detector material. Finally, in the Appendix B, a discussion on the modeling of plasmonic gold as an effective sheet impedance in simulation is provided.

An $e^{j\omega t}$ time convention is assumed and suppressed throughout.

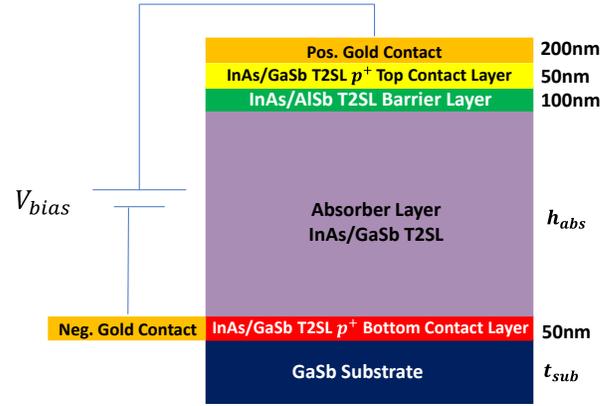

Fig 2. pBp infrared photodiode stackup. The active region is an InAs/GaSb T2SL. The highly doped contact layers are also InAs/GaSb T2SL layers. The wide bandgap barrier layer is an InAs/AlSb T2SL.

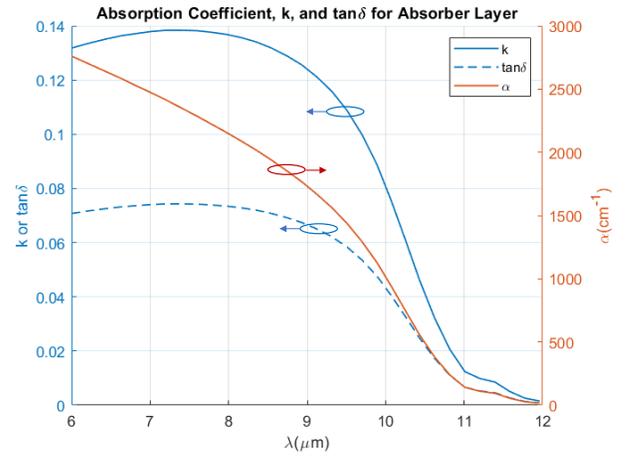

Fig 3. Absorber layer effective material parameters. The data marked $\alpha$ is modeled from that found in [16].

## II. ANTIMONIDE PHOTODETECTOR DESIGN

Previously, we reported measured results of a DRA coupled to a bulk InAs$_{1-x}$Sb$_x$ infrared photodetector whose composition was tuned to achieve a bandgap of $9\mu$m [13]. This device was designed as a frontside illuminated geometry with a highly doped ground plane to improve coupling to the antenna from the surrounding air. However, for imaging applications using focal plane arrays, the light is incident through the substrate. We explore a backside illuminated design in this paper that further improves the absorption of a small detector element in a way that is compatible with conventional hybridization techniques for integration with read-out integrated circuits (ROICs). Samples of InAs/GaSb strained layer superlattices have been grown and extraction of the optical properties was performed [14,15].

The photodetector consists of a pBp photodiode architecture using an InAs/GaSb Type-II superlattice (T2SL) absorber active region with an $11\mu$m cutoff and highly doped contact regions as shown in Fig. 2. The barrier is a wide band gap InAs/AlSb T2SL. Superlattice materials typically have a lower absorption coefficient than their bulk counterparts and thus will benefit more from the added DRA. Ohmic contacts are established with gold structures on top of the device and along the bottom $p^+$ ground plane.

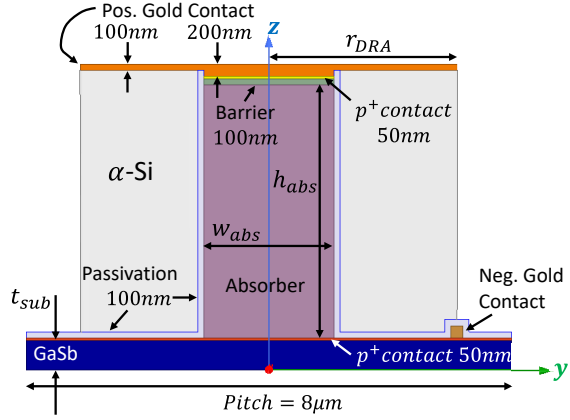

Fig 4. DRACAD Geometry. The photodetector mesa sits atop a GaSb substrate. A Al2O3 passivation layer blankets the substrate and detector mesa. An amorphous Si DRA is deposited over the passivation layer. The DRACAD is topped with a gold contact.

The pBp detector mesa was homogenized and modeled with effective material parameters from literature [16-18]. Table I provides the material parameters and dimensions for the various layers. All four layers ($p^+$ top contact, barrier, absorber, and $p^+$ bottom contact) were assigned an index of refraction of 3.73. These layers were assumed lossless (as their thicknesses are very thin compared to the target wavelength) apart from the absorber layer whose absorption coefficient is provided in Fig. 3 based on parameters from [16]. Detector parameters that were optimized during the simulation process are listed as variables in Table I.

TABLE I
DRACAD GEOMETRY AND MATERIAL PARAMETERS

| Layer | Thickness | Width | n | k |
|---|---|---|---|---|
| GaSb Substrate | $t_{sub}$ | 8µm | See [19] | See [19] |
| $p^+$ Bottom Contact | 50nm | 8µm | 3.73 | 0 |
| Absorber | $h_{abs}$ | $w_{abs}$ | 3.73 | See Fig. 3 |
| Barrier | 100nm | $w_{abs}$ | 3.73 | 0 |
| $p^+$ Top Contact | 50nm | $w_{abs}$ | 3.73 | 0 |
| DRA | $h_{abs}$+150nm | $r_{DRA}$ | 3.27 | 0 |
| Passivation | 100nm | N/A | See [20] | See [20] |
| Pos. Gold Contact | 100nm/200nm | $r_{DRA}$ | See Ap.B | See Ap.B |
| Neg. Gold Contact | 100nm | 100nm | See Ap.B | See Ap.B |

Variable quantities entered in the dimension columns are optimized variables. The optimized values are provided in section III.B

## III. DIELECTRIC RESONATOR ANTENNA DESIGN

### A. Design of Dielectric Resonator Antenna Coupled Detector

The Dielectric Resonator Antenna Coupled Antimonide Detector (DRACAD) geometry is shown in Fig. 4. The relevant dimensions and material properties are provided in Table I. A GaSb substrate supports the infrared photodetector mesa described in section II. The mesa is centered within an 8µm x 8µm unit cell of the Focal Plane Array (FPA). The $p^+$ highly doped contact layer sits atop the GaSb substrate. A gold contact is positioned 3µm to the right of the center of the unit cell and is in contact with the $p^+$ contact layer. This gold bar runs the full depth of the unit cell and serves as the negative D.C. ohmic contact used to extract the current from and bias the detector.

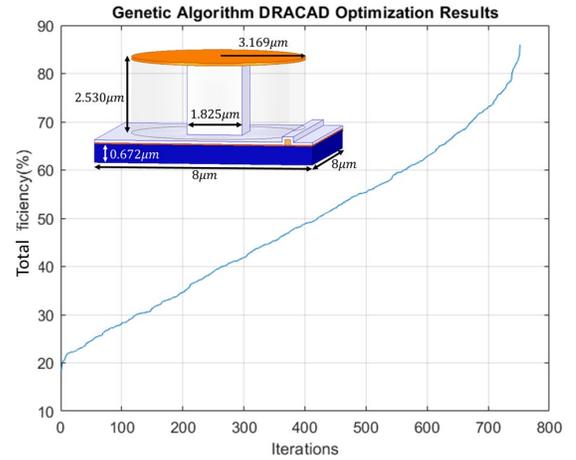

Fig 5. Optimization results for the DRACAD design. The 3D model of the optimized geometry is shown in the inset of the figure.

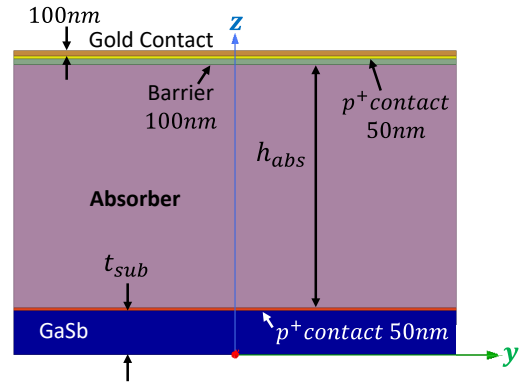

Fig 6. Conventional resonant cavity enhanced slab detector geometry. The conventional resonant cavity enhanced slab detector is infinite in the $x$ and $y$ directions and has an optimized resonant thickness.

The gold bar has dimensions of 100nm x 100nm x 8000nm. A 100nm thick Al2O3 passivation layer blankets the $p^+$ layer, the negative gold contact, and detector mesa sidewalls. The amorphous Silicon (α-Si) DRA is deposited over the passivation layer. The DRA surrounds the detector and forms a resonant cavity (antenna) with the detector acting as a load. The DRA is capped with a gold layer, which serves as the positive ohmic contact. The DRACAD was designed for backside illumination. Thus, the excitation is a plane wave incident from below the substrate. The unknown geometrical parameters appearing in Table I are determined through optimization.

### B. Optimization of Dielectric Resonator Antenna Coupled Detector

The role of the DRA is to both provide a large effective aperture area for the small detector and to impedance match the incident radiation to the detector impedance. The impedance matching is important to minimize reflection at the air-dielectric interface, and thus couple in as much light as possible. Thus, optimal geometrical parameters should be found that serve both purposes.

The variables defining the DRACAD geometry ($t_{sub}, w_{abs}, h_{abs}, r_{DRA}$) are optimized using a Genetic Algorithm (GA). These geometrical parameters are optimized for maximum total efficiency at 9µm with minimum detector volume. The total efficiency is defined in (A.8) as $P_{abs}/P_{inc}$ where $P_{abs}$ is the power absorbed in the detector volume and $P_{inc}$ is the power in the incident plane wave.

In the simulation, the sidewalls of the simulation domain are periodic. The top surface of the simulation domain is covered by a Perfectly Matched Layer (PML). The bottom surface is selected as a Floquet port with a single Floquet mode incident. This Floquet mode (along with the degenerate orthogonally polarized mode) is the only propagating mode for normally incident radiation when the array spacing is $8\mu m$ and the operating wavelength is $9\mu m$. The GA fitness function to maximize was set to

$$fitness = 0.8\eta + 0.2\left(1 - \frac{V_{abs}}{V_{DRA}}\right) \quad (1)$$

The weighted first term maximizes the total efficiency, $\eta$ (or $P_{abs}/P_{inc}$) in the pBp photodetector absorber region while the second weighted term minimizes the volume of the detector, $V_{abs}=w_{abs}^2 h_{abs}$, with respect to the volume of the DRA antenna, $V_{DRA}=\pi r_{DRA}^2 h_{abs}$. The fitness function is designed to obtain maximum absorption with minimum detector dimensions thereby improving the signal received and the NEP. Consequently, the optimizer also determines the correct dimensions to impedance match the incident field to the detector impedance. Since the incident Floquet mode is normalized to $P_{inc}=1W$ power over the unit cell area, the total efficiency is calculated using

$$\eta = \frac{1}{2}\text{Re}\left[\int_{-\frac{w_{abs}}{2}}^{\frac{w_{abs}}{2}}\int_{-\frac{w_{abs}}{2}}^{\frac{w_{abs}}{2}}\int_{-\frac{h_{abs}}{2}}^{\frac{h_{abs}}{2}} \sigma |\vec{E}|^2 dxdydz\right] \quad (2)$$

where the limits of integration are over the volume of the absorber region only. The effective conductivity appearing in (2) is calculated from the parameters in Table I as

$$\sigma(\omega) = j\varepsilon_0\omega\left(n^2 - k^2 - 1 - j2nk\right) \quad (3)$$

The optimizer was run for approximately 750 iterations and converged upon a solution which absorbs 86% of the incident light as shown in Fig. 5. The converged design parameters are shown in the inset of Fig. 5 and provided in Table II.

TABLE II
OPTIMIZED DRACAD GEOMETRY PARAMETERS

| PARAMETER | $t_{sub}$ | $w_{abs}$ | $h_{abs}$ | $r_{DRA}$ |
|---|---|---|---|---|
| VALUE | $0.672\mu m$ | $1.825\mu m$ | $2.530\mu m$ | $3.169\mu m$ |

The optimized fitness was 0.860 (the first term of (1) contributed 0.681 and the second term contributed 0.179).

For comparison purposes, a conventional resonant cavity enhanced slab detector was optimized for maximum absorption at the target wavelength of $9\mu m$ using the same technique [11,12]. The optimization allows the conventional detector geometry to impedance match the incident radiation to the absorbing slab. If an infinite half space were used to collect the radiation rather than a slab with optimized resonant thickness, the maximum total efficiency achievable would be limited by the impedance mismatch at the air-detector interface. The maximum total efficiency for a half space is $\eta = (1 - r) = 0.68$ (68%), where $r$ is the reflectivity of the half space made from absorber material. Optimizing the thickness of the conventional detector can exceed this maximum.

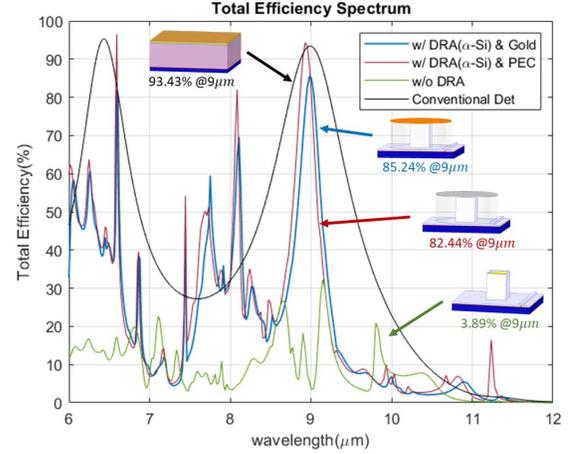

Fig. 7. Total Efficiency Spectra. Shown are four curves: the optimized DRACAD with PEC in place of all metals (w/ DRA($\alpha$-$Si$) & PEC), the optimized DRACAD with gold metals (w/ DRA($\alpha$-$Si$) & Gold), the optimized DRACAD with DRA removed (w/o DRA), and the optimized conventional detector (Conventional Det).

The conventional resonant cavity enhanced slab detector stackup to optimize, shown in Fig. 6, consists of: the GaSb substrate, the $p^+$ bottom contact layer, the InAs/GaSb T2SL absorbing layer, the InAs/AlSb barrier layer, the $p^+$ top contact layer, and a Gold ohmic contact. The optimization variables are the thickness of the substrate, $t_{sub}$, and the thickness of the absorber layer, $h_{abs}$. The fitness function was similar to (1) except only the first term is kept. The optimized geometrical parameters are shown in Table III. Note that the conventional detector and the DRACAD have nearly identical absorber layer thicknesses. This makes their NEP comparisons more direct. Their volumes differ, however, by 95%.

TABLE III
OPTIMIZED CONVENTIONAL RESONANT CAVITY ENHANCED SLAB DETECTOR GEOMETRY PARAMETERS

| PARAMETER | $t_{sub}$ | $h_{abs}$ |
|---|---|---|
| VALUE | $0.373\mu m$ | $2.419\mu m$ |

### C. Total Efficiency Spectra

The total efficiency spectrum of the optimized geometry of the DRACAD was computed next. The excitation frequency of a normally incident plane wave from the below the GaSb substrate was varied throughout the LWIR band of wavelength range of 6-12$\mu$m. The total efficiency was calculated using (2) . The number of Floquet modes was increased to 10. These 10 modes are the only propagating modes for the shortest wavelength of operation ($\lambda_0 = 6\mu m$) and the incidence is kept normal. The total efficiency was calculated for various cases: **1.** for the optimized DRACAD, **2.** for the optimized DRACAD with Perfect Electric Conductor (PEC) in place of the gold material, **3.** with the DRA removed from the optimized DRACAD in order to show the DRA's effect on total efficiency, and **4.** the optimized conventional resonant slab detector of Fig. 6. The spectra are shown in Fig. 7. The total efficiencies associated with each case in Fig. 7 are displayed in

Table IV.

TABLE IV
TOTAL EFFICIENCIES OF VARIOUS ABSORPTION SPECTRA CASES

| Case | $\eta(9\mu m)$ | $\lambda_{abs,peak}$ | $\eta(\lambda_{abs,peak})$ |
|---|---|---|---|
| **1:** w/ DRA ($\alpha$-Si) & Gold | 85.24% | 9.000$\mu m$ | 85.24% |
| **2:** w/ DRA ($\alpha$-Si) & PEC | 82.44% | 8.929$\mu m$ | 94.34% |
| **3:** w/o DRA | 3.89% | 9.146$\mu m$ | 32.29% |
| **4:** Conventional Detector | 93.43% | 9.000$\mu m$ | 93.43% |

The bold number indicates the case number.
The shaded grey row in the table is the final DRACAD design.

The table shows the DRACAD geometry (Case 1) provides a maximum total efficiency of 85.24%. For comparison, in [5], the maximum absorption in the active layer (absorbing layer) is 47%, and in [6], the maximum is 52.8%.

By replacing the lossy gold with PEC (Case 2), the total efficiency absorption peak wavelength, $\lambda_{abs,peak}$, has shifted to 8.929$\mu m$ since the inductive reactance of the gold material was removed. The loss due to the plasmonic gold can be estimated by subtracting the total efficiencies at the absorption peak wavelength as 0.091 or 9.1%.

By removing the DRA from the DRACAD (Case 3) the $\eta$ has reduced to 3.89% at 9$\mu m$. This clearly demonstrates the impedance matching the DRA provides and the resonant cavity field enhancement effect of the DRA on the $\eta$.

The conventional detector's total efficiency at 9$\mu m$ is 93.43% which is 8.19% higher than the DRACAD total efficiency at the same wavelength. Thus, the DRACAD collects nearly the same amount of signal as the conventional detector. The DRACAD however, has superior noise performance, as will be shown in the next section.

## IV. SIGNAL TO NOISE RATIO

In this section, the SNR will be calculated and compared for both the conventional detector and the DRACAD. The noise power can be calculated from (A.13) and (A.14). The dark current appearing in (A.13) and (A.14), $I_{dark}$, can be extracted from the measurements provided in [16]. At 77K, the dark current for the 8$\mu m$ × 8$\mu m$ sized unit cell of the DRACAD is $I_{dark} = 2.24 \times 10^{-11}A$. From (A.13), the NEP for the DRACAD can be determined. From (A.14) the NEP for the conventional detector can be determined. The signal power is estimated as the power illuminating an 8$\mu m$ × 8$\mu m$ unit cell from a 300K black body emitter of emissivity 1 and within the wavelength range of 8.99-9.01$\mu m$. Assuming a $f/2$ optics system, this signal power is $P_{inc} = 3.3278 \times 10^{-13}$ W. Over this wavelength range, $\eta$ is nearly constant and equal to the values quoted in Table IV. The SNR for each case is then $SNR = P_{inc}/NEP$. The NEP and SNR for both cases are provided in Table V.

TABLE V
NOISE EQUIVALENT POWER AND SIGNAL-TO-NOISE RATIO COMPARISON

|  | $NEP$ (W) | $P_{inc}$ (W) | $SNR$ (dB) |
|---|---|---|---|
| Conv. Detector | $4.2313 \times 10^{-16}$ | $3.3278 \times 10^{-13}$ | 28.96 |
| DRACAD | $1.0580 \times 10^{-16}$ | $3.3278 \times 10^{-13}$ | 34.98 |

The NEP values in the table assume a 1Hz noise bandwidth.

The NEP calculations appearing in Table V assume a 1Hz noise bandwidth and thus can be scaled to any appropriate noise bandwidth by multiplication by a factor of $\sqrt{\Delta f}$. Note, the pBp photodiode structure and the small pixel cross sectional area lead to low dark currents and therefore NEP [16]. The table shows the DRACAD improves the NEP by a factor of 4 and the SNR by 6.02dB. The DRACAD collects nearly the same signal as the conventional detector while generating less noise. Additionally, the DRACAD inherently has 8$\mu m$ × 8$\mu m$ resolution in the FPA, and each of the 8$\mu m$ × 8$\mu m$ pixels in the FPA have a 6.02dB advantage over the conventional resonant cavity enhanced slab detector.

## V. HIGH RESOLUTION FOCAL PLANE ARRAYS: COMPARISON TO STATE OF THE ART

A comparable technology which utilizes resonant antennas to concentrate incident radiation onto reduced dimension detectors are plasmonic antenna coupled infrared photodetectors [21]. The antennas made from plasmonic metals compresses the wavelength and focuses the energy onto a small detector. In [21], dipole, bowtie, spiral, and log periodic plasmonic antennas were coupled to reduced dimension bolometers for detection at 10.6$\mu m$. The reference reports calculated radiation efficiencies of 20%, 37%, 25%, and 46%, respectively. In [22], plasmonic nanopillar optical antennas are coupled to nBn infrared photodetectors. The nanopillar optical antennas are photon-trapping devices which concentrate the incident radiation onto the small pixels. The reference reports a total efficiency of only 51%. The DRACAD achieves much higher efficiencies as the metallic losses are avoided by choosing an all dielectric resonant antenna as the concentrator.

A better comparison may then be infrared detectors coupled to all dielectric concentrators such as a microlens array. In microlens arrays, dielectric lenses are placed above each pixel in a focal plane array. The microlens concentrates the incident light onto a reduced dimension photodetector. In [23], an array of Si hemispherical microlenses was grown over thermopile infrared photosensors for operation over a wavelength range of 8-14$\mu m$. Signal collection was enhanced by 2.34 times due to the microlens array. Similar enhancements were reported in [24] where an array of Fresnel zoned microlenses was fabricated over an array of photosensors. Over the wavelength range of 8$\mu m$-12$\mu m$, [24] reports collecting over 80% of the incident light resulting in a signal collection enhancement factor of 2.5 times that without the microlenses present. In comparison, the DRACAD collects over 85% of the incident light and enhances signal generation by 2.64 times if the peak wavelengths in Table IV are used. Note, the factor of 2.64 comes from forming the ratio of $\eta(\lambda_{abs,peak})$ for case **1** and case **3**. However, the microlens array's elements are much larger than those of the DRACAD. In particular, the microlens array unit cells in [23] occupy a cross-sectional area of 500$\mu m^2$. In contrast, the DRACADs unit cells occupy a cross-sectional area of 64$\mu m^2$. Also, the DRACADs unit cell volume is 0.000656 (0.066%) that of the microlens array unit cells (the thickness of the microlens array is approx. 200 times thicker than the DRACAD). The FPA resolution of the DRACAD is 0.79 square wavelengths, whereas the microlens array is 6.17244 square wavelengths. *N*=8 DRACAD elements can fit in the same area as a single microlens, and because they have nearly the same signal enhancement factor, the received signal of the combined DRACADs is then increased by a factor of $N$=8 while the noise only increases by a factor of $\sqrt{N}$ leading to an enhanced SNR. These eight-parallel connected DRACADs total photodetector volume is 26.6$\mu$m³. The single microlens array detector has a volume greater than 100$\mu$m³. Thus, the active material volume inherent in the microlens array is much greater than the DRACAD which will lead to more noise generated by the microlens array system.

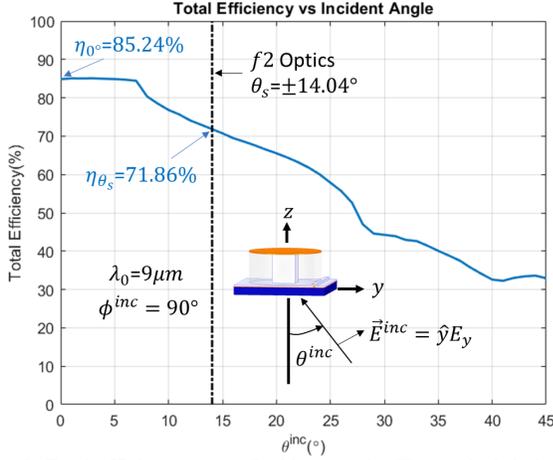

Fig. 8. Total efficiency versus Incident Angle. The vertical dashed line indicates the subtended angle for $f2$ optics. The total efficiency is relatively constant over the incident angles corresponding to the illuminating $f2$ optics.

The DRACADs performance is far superior due to its small footprint: better FPA resolution, high pixel count cameras, lower noise, higher received signal, and smaller volume. One key factor of the DRACAD is its focal plane array resolution. Trends in state-of-the-art infrared focal plane arrays is the push to higher pixel count cameras and better focal plane array resolution [25]. The DRACAD is expected to be the state-of-the-art in future high pixel count, high resolution, low noise, focal plane arrays in the future.

## VI. INCIDENT ANGLE PERFORMANCE

Let us assume that the DRACAD design will be placed in a focal plane array with $f/2$ optics. Thus, the subtended angle is $\theta_s$=14.04°. In Fig. 8, the total efficiency was simulated versus incident angle. At this subtended angle, the total efficiency $\eta_{\theta_s}$ is 71.86% which is a drop of only 13.38% from the on-axis value as shown in Fig. 8. Since the optimum DRACAD geometry maximizes the impedance match between the incident wave impedance and the detector load impedance, any change in the incident angle will reduce the degree of matching resulting in a decrease in total efficiency and thus SNR. Figure 8 shows that the optimized DRACAD's total efficiency remains relatively constant over the $f/2$ subtended angle range and is compatible with $f/2$ optics.

## VII. CONCLUSION

A dielectric resonator antenna coupled infrared photodetector is introduced. The antenna coupled detector operates in the backside illuminated mode that is compatible with conventional hybridization techniques for integration with read-out integrated circuits (ROICs). The detector is a pBp photodiode with an InAs/GaSb Type-II superlattice active region. The dielectric resonator antenna and detector dimensions were optimized using a Genetic Algorithm. The optimizer maximized the absorbed photons while simultaneously minimizing the detector volume. Since noise power is generated throughout the volume of the detector, the small detector volume reduces the noise power. The dielectric resonator collects over a wide aperture and concentrates the power onto the small detector and impedance matches the detector to the incident wave impedance. It was found that the dielectric resonator antenna coupled detector outperformed the conventional detector made from a resonant cavity enhanced slab by 6.02dB in SNR per 8$\mu$m × 8$\mu$m unit cell. The noise power was calculated from derived formulas involving the Noise Equivalent Power (NEP). The signal power was obtained through full wave simulations of the optimized antenna coupled detector. Plasmonic gold materials were modeled as effective sheet impedances. Simulations of total efficiency (product of the antenna aperture efficiency and the quantum efficiency of the photodetector) versus incident angle shows the antenna coupled detector has nearly constant total efficiency for all angles associated with $f/2$ optics. It is expected that FPA's constructed with DRACAD elements will provide enhanced SNR and superior resolution than traditional infrared photodetectors and allow for higher operating temperatures.

## APPENDIX A. THEORETICAL OVERVIEW OF ANTENNA COUPLED PHOTODETECTORS

Infrared photodetectors convert *internal* infrared radiation power $P$ into an output voltage $V_s$ or current $I_s$. The proportionality constant which relates this input power to the output voltage or current defines the photodiodes transfer function and is called the Responsivity ($R$). The Responsivity therefore has units of Volts/Watt for voltage responsivity ($R_v$) or Amps/Watt for current responsivity ($R_i$). The input power that generates an output voltage (current) equal to the RMS voltage (current) of the noise generated by the device is referred to as the Noise Equivalent Power. Thus, the NEP is the input power which produces an output Signal to Noise Ratio (SNR) of 1. In terms of $R_v$ ($R_i$), the NEP can be written as

$$NEP_{\text{int}} = \frac{V_n}{R_v} = \frac{I_n}{R_i} \qquad (A.1)$$

where $V_n$ is the RMS noise voltage and $I_n$ is the RMS noise current. The lower this number is, the lower the noise floor is and the larger the R is of the photodetector. One major contribution of this paper is to reduce the NEP by coupling an all-dielectric antenna to the photodetector. The antenna coupled configuration improves the SNR at the output. The increase in the SNR will be quantified through theoretical calculations and simulation results. An analytical expression relating the improvement in the SNR to the dimensions of the antenna and detector geometry is derived next.

An equivalent figure of merit to the Noise Equivalent Power (NEP), is the specific Detectivity ($D^*$) [10]. The NEP is proportional to the square root of the detector signal, which in turn is proportional to the detector area $A_d$ and the noise bandwidth [26]. The NEP can then be defined as the reciprocal of a detector dimension and bandwidth independent constant ($D^*$) scaled by the square root of bandwidth and detector area as [27]

$$NEP_{int} = \frac{(A_d \Delta f)^{1/2}}{D^*} \quad (A.2)$$

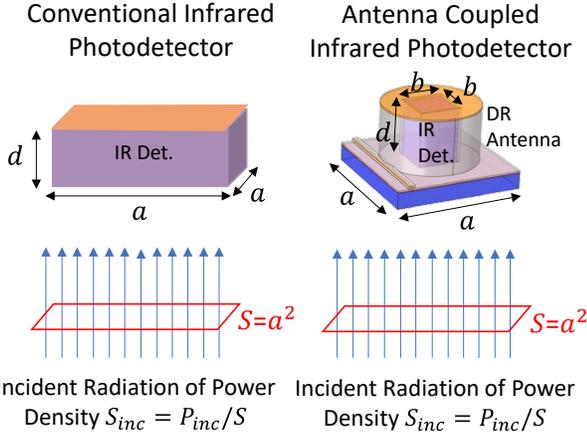

Fig A.1. Antenna coupled infrared photodetector concept. The conventional infrared photodetector is a slab of cross sectional area $a \times a$ and resonant thickness with incident power $P_{inc}$ over an aperture area of S (Poynting vector $\vec{S}^{inc} = \hat{z}S^{inc}$). The antenna coupled infrared photodetector is a reduced dimension infrared photodetector of cross sectional area $b \times b$ loaded resonant cavity antenna which also has incident power $P_{inc}$ illuminating an aperture area of S (Poynting vector $\vec{S}^{inc} = \hat{z}S^{inc}$).

The detector dimension and bandwidth independent constant is called the Specific Detectivity ($D^*$). The advantage of $D^*$ as a figure of merit for infrared detectors is it allows for the comparison of detectors of the same material type but of different sizes and operational bandwidth. This makes the figure independent of application specific details. The importance of (A.2) is that it shows the NEP is proportional to the square root of detector area $A_d$. Thus, if the dimensions of the detector are reduced, then the NEP is also reduced.

The antenna coupled detector NEP will be compared to the NEP of a conventional resonant slab detector made from a planar infinite slab made of absorbing material of resonant thickness. The comparison is depicted in Fig. A.1. The specific detectivity $D^*$ of the conventional detector and that of the antenna coupled detector is the same since the active region of each is constructed from the same absorbing material. Their NEP's, however, are different given that the antenna coupled detectors absorbing active region is much smaller. The NEP of the conventional detector is derived first. Then the NEP for the antenna coupled detector is derived and the ratio taken to obtain an expression for the improvement in NEP due to the antenna coupled configuration.

The power coupled into a conventional detector from external infrared radiation is $P_L = P_{inc}\eta_{ap} = P_{inc}\sigma_{conv}/a^2$, where $P_{inc}$ is the incident power in Watts over pixel aperture area $a^2$, $\eta_{ap}$ is the aperture efficiency of the conventional detector, and $\sigma_{conv}$ is the effective area in square meters of the conventional detector. The antenna aperture efficiency is also known as the *external* quantum efficiency. Note, by definition, the effective area accounts for ohmic loss and reflection loss due to the impedance mismatch at the interface. With this definition of power coupled into the detector, (A.1) can be re-expressed in the language of an *external* Responsivity as

$$\frac{P\sigma_{conv}}{a^2} = \frac{I_n}{R_i} \rightarrow NEP_{ext} = P = \left(\frac{a^2}{\sigma_{conv}}\right)\left(\frac{I_n}{R_i}\right) \quad (A.3)$$

It is important to note that (A.3) differs from (A.1) in that the input power is referenced outside of the device to the illuminating wave rather than the power already coupled into the detector. By equating (A.2) and (A.3), an expression can be derived for the NEP in terms of the detector dimensions of the conventional resonant cavity enhanced slab of physical cross-sectional area $a \times a$

$$\frac{P\sigma_{conv}}{a^2} = \frac{I_n}{R_i} = \frac{\sqrt{A_d \Delta f}}{D^*} \rightarrow$$

$$(NEP_{ext})_{conv} = P = \frac{a^2 a}{\sigma_{conv} D^*}\sqrt{\Delta f} \quad (A.4)$$

In the antenna coupled configuration, the detector has cross-sectional area $b \times b$ rather than $a \times a$, and is coupled to an antenna with effective area $\sigma_{ant}$ square meters. The amount of power delivered to the load by the antenna is therefore $P_L = P_{inc}\eta_{ap} = P_{inc}\sigma_{ant}/a^2$, where $P_{inc}$ is the incident power in Watts over pixel aperture area $a^2$ and $\eta_{ap}$ is the aperture efficiency of the antenna. The NEP of the antenna coupled configuration is

$$\frac{I_n}{R_i} = \frac{P\sigma_{ant}}{a^2} = \frac{\sqrt{A_d \Delta f}}{D^*} \rightarrow$$

$$(NEP_{ext})_{ant} = P = \frac{a^2 b}{\sigma_{ant} D^*}\sqrt{\Delta f} \quad (A.5)$$

The NEP improvement factor for the antenna coupled configuration is found by ratioing the two cases

$$(NEP_{ext})_{ant} = (NEP_{ext})_{conv}\left(\frac{b\sigma_{conv}}{a\sigma_{ant}}\right) \quad (A.6)$$

If each photon absorbed in the absorbing medium generates a photocarrier and each photocarrier is collected, the number of photocarriers generated per number of incident photons defines the *internal* quantum efficiency, $\eta_{qe}$, as the ratio between the absorbed power $P_{abs}$ to the power delivered to the detector load, $P_L$ [29]

$$\eta_{qe} = \frac{P_{abs}}{P_L} \quad (A.7)$$

Note, due to the assumption above (all photons absorbed will generate carriers which are collected), the quantum efficiency will be unity, $\eta_{qe}=1$. Assuming the antenna is impedance matched to the detector, the total efficiency is found from

$$\frac{P_{abs}}{P_{inc}} = \eta_{ap}\eta_{qe} = \eta \quad (A.8)$$

Thus, the effective areas can be defined in terms of the total efficiencies as

$$\sigma = \frac{P_{abs}}{S_{inc}} = \frac{\eta P_{inc}}{S_{inc}} = \frac{\eta(1W)}{(1W/a^2)} = \eta a^2 \quad (A.9)$$

where $P_{abs}$ is the received power absorbed by the load (absorbed by the absorber), $P_{inc}$ is the incident power, $S_{inc}$ is the incident field power density, and $\eta$ is the respective total efficiency for either the conventional detector, $\eta_{conv}$, or the antenna coupled detector, $\eta_{ant}$. With (A.9), (A.6) can be written as

$$\frac{(NEP_{ext})_{ant}}{(NEP_{ext})_{conv}} = \left(\frac{b\eta_{conv}}{a\eta_{ant}}\right) \quad (A.10)$$

Therefore, improvement in NEP is proportional to the ratio of $b/a$. The total efficiencies $\eta_{conv}$ and $\eta_{ant}$ can be obtained through simulations.

Using the definition of $(NEP_{ext})_{conv}$ from (A.3) and the effective area from (A.9), (A.10) becomes

$$(NEP_{ext})_{ant} = \left(\frac{1}{\eta_{conv}}\right)\left(\frac{I_n}{R_i}\right)\left(\frac{b\eta_{conv}}{a\eta_{ant}}\right)$$
$$= \left(\frac{1}{\eta_{ant}}\right)\left(\frac{\sqrt{2qI_{dark}\Delta f}}{R_i}\right)\left(\frac{b}{a}\right) \quad (A.11)$$

where $I_n = \sqrt{2qI_{dark}\Delta f}$ [27, p.130]. $I_{dark}$ is the D.C. dark current for the conventional resonant slab photodetector, $q$ is the electron charge, and $\Delta f$ is the noise bandwidth. The Responsivity can be defined in terms of the total efficiency by

$$R_i = \frac{q\lambda}{hc}\eta \quad (A.12)$$

where $\lambda$ the operating wavelength, $h$ is plank's constant, and $c$ is the speed of light. From (A.12), (A.11) can be expressed in terms of the total efficiency and the dark current as

$$(NEP_{ext})_{ant} = \left(\frac{b}{a}\right)\left(\frac{1}{\eta_{ant}}\right)\left(\frac{\sqrt{2}hc}{\sqrt{q}\lambda}\right)\left(\frac{\sqrt{\Delta f}\sqrt{I_{dark}}}{\eta_{conv}}\right)$$
(A.13)

Thus, the NEP of the antenna coupled detector can be calculated from knowledge of the total efficiency of the conventional resonant slab detector, $\eta_{conv}$, the total efficiency of the antenna coupled detector, $\eta_{ant}$, and the dark current of the conventional resonant cavity enhanced photodetector, $I_{dark}$. All three parameters are determined in this work. Following from (A.10), the NEP of the conventional resonant cavity enhanced detector can be expressed in similar terms as

$$(NEP_{ext})_{conv} = \left(\frac{1}{\eta_{conv}}\right)\left(\frac{\sqrt{2}hc}{\sqrt{q}\lambda}\right)\left(\frac{\sqrt{\Delta f}\sqrt{I_{dark}}}{\eta_{conv}}\right)$$
(A.14)

These definitions (A.12 and A.13) are used to quantify improvements to the SNR of the antenna coupled design.

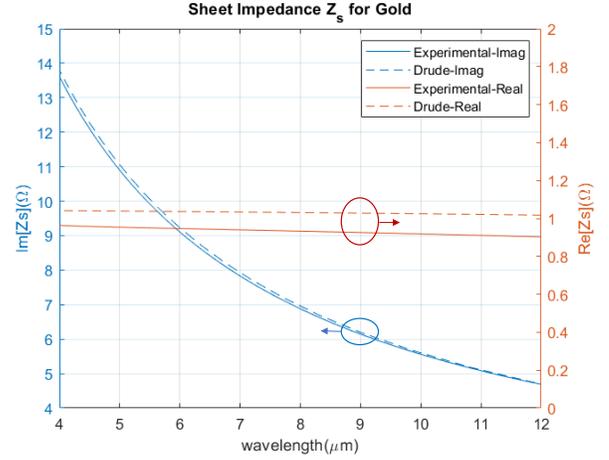

Fig B.1. Effective sheet impedance for Gold material. The dashed curves are from the Drude model. The solid curves are from experimental data taken from [28]. The effective surface impedance is inductive and lossy.

APPENDIX B. MODELING OF PLASMONIC GOLD AS SURFACE IMPEDANCE

At the THz frequencies of operation, gold is plasmonic. The Drude model can be used to model its complex dielectric function [25]

$$\varepsilon_r(\omega) = 1 - \frac{\omega_p^2}{\omega^2 - j\Gamma\omega} \quad (B.1)$$

where $\omega$ is the angular frequency of excitation, $\omega_p = 1.2914 \times 10^{16}$ rad/sec is the plasma frequency, and $\Gamma = 1/\tau_D = 7.1429 \times 10^{13}$ sec$^{-1}$ is the electron relaxation rate. The complex conductivity can be obtained from (B.1)

$$\varepsilon_r' = \Re[\varepsilon_r(\omega)]$$
$$\varepsilon_r'' = \Im[\varepsilon_r(\omega)] \quad (B.2)$$
$$\sigma(\omega) = j\varepsilon_0\omega(\varepsilon_r' - j\varepsilon_r'' - 1)$$

where $\epsilon_0 = 8.854 \times 10^{-12}$ F/m is the permittivity of free space. Note, $\Re$ denotes the real part and $\Im$ denotes the imaginary part. The effective surface impedance is then found as [28]

$$Z_s(\omega) = \sqrt{\frac{j\omega\mu_0\mu_r}{\sigma(\omega) + j\omega\varepsilon_0}} \quad (B.3)$$

where $\mu_0 = 4\pi \times 10^{-7}$ H/m is the permeability of free space. The relative permeability, $\mu_r$, of gold is equal to 1. The surface impedance (B.3) is plotted in Fig. B.1. The figure shows both the real part (resistance) and imaginary part (inductive reactance) for the gold material over the LWIR wavelength range. Also plotted in Fig. B.1 is the experimental data taken in [28]. The Drude model shows good agreement with measurements.

In order to validate the sheet impedance model for the gold material, the power absorbed by an infinite gold slab modeled with Drude dispersion was compared to that from an infinite sheet with surface impedance given by (B.3). The slab thickness is increased in steps of skin depth $\delta$ until agreement in the absorbed power is obtained. The skin depth $\delta$ is found to be

$$\delta(\omega) = \frac{c}{2\omega k} = \frac{c}{2\omega \Im\left[\sqrt{\varepsilon_r'(\omega) - j\varepsilon_r''(\omega)}\right]} \quad (B.4)$$

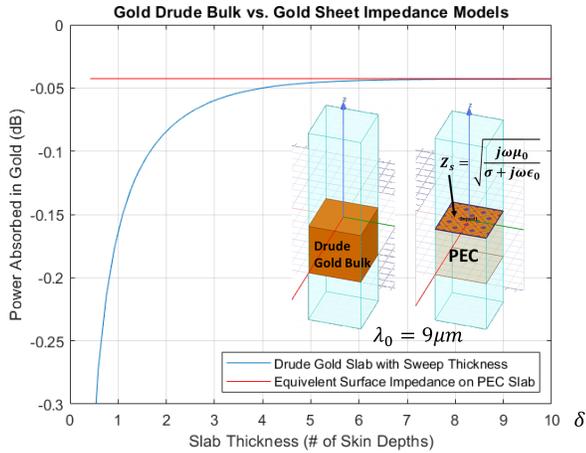

Fig B.2. Skin depth test for sheet impedance model of Gold material. The 'Drude gold slab with Sweep Thickness' curve results from sweeping the thickness of an infinite planar slab of bulk gold material modeled with the Drude dispersion and calculating the absorbed power for each thickness. The Power absorbed in the gold is calculated by determining the difference between the incident power and the power received at each of the ports located above and below the slab. The 'Equivalent Surface Impedance on PEC Slab' curve results from calculating the difference between the incident power and the power received at the port located above the slab.

The skin depth from (B.4) at $\lambda_0 = 9\mu m$ is 11.95nm. A comparison of the power absorbed by the infinite slab and sheet impedance is shown in Fig. B.2. The equivalent surface impedance model is shown to be accurate for gold layer thicknesses of at least 7 skin depths or 83.65nm. Since each of the gold layers in Fig. 2 and Fig. 4 are at least 100nm, the effective surface impedance model is valid. Thus, the loss associated with dissipation in the bulk gold is fully modeled by the surface loss associated with the effective surface impedance. The advantage of using the surface impedance model is that only the surface is meshed in simulation rather than the bulk or volume. Since the skin depth in the gold material is 11.95nm, the fields within the gold material decay rapidly. In order to resolve the fields within the gold material, extremely fine volumetric discretization would be required in simulation. By utilizing the effective surface impedance model, the computational resource requirements for the simulation are reduced making optimizations possible.